\documentclass[12pt,preprint]{aastex}
\usepackage{lscape}

\def\e{{\rm E}}
\def\au{{\rm AU}} 
\def\muas{{\mu\rm as}}
\def\kms{{\rm km}\,{\rm s}^{-1}}
\def\kpc{{\rm kpc}}
\def\cm{{\rm cm}}
\def\pc{{\rm pc}}
\def\mpc{{\rm Mpc}}

\def\e{{\rm E}}
\def\ray{{\rm ray}}
\def\vecr{{\bf r}}
\def\vecmu{\mbox{\boldmath$\mu$}}
\def\vecv{{\bf v}}
\def\abs{{\rm abs}}
\def\vectheta{\mbox{\boldmath$\theta$}}

\begin{document}
\title{Probing $\sim 100 \au$ Intergalactic \ion{Mg}{2} Absorbing
``Cloudlets'' with Quasar Microlensing}

\author{Subo Dong}
\affil{
Department of Astronomy, The Ohio State University,
140 W.\ 18th Ave., Columbus, OH 43210, USA
}
\email{dong@astronomy.ohio-state.edu}

\begin{abstract}
Intergalactic \ion{Mg}{2} absorbers are known to have structures down
to scales $\sim 10^{2.5} \pc$, and there are now indications that they
may be fragmented on scales $\lesssim 10^{-2.5} \pc$ (Hao et al.,
astro-ph/0612409). 
When a lensed quasar is microlensed, the micro-images of the 
quasar experience creation, destruction, distortion, and drastic
astrometric changes during caustic-crossing.
I show that quasar microlensing can effectively probe
\ion{Mg}{2} and other absorption ``cloudlets'' with sizes $\sim
10^{-4.0} - 10^{-2.0} \pc$ by inducing significant spectral
variability on the timescales of months to years. With numerical
simulations, I demonstrate the feasibility of applying this method to
Q2237+0305, and I show that high-resolution spectra of this quasar in
the near future would provide a clear test of the existence of such
metal-line absorption ``cloudlets'' along the quasar sight line.

\end{abstract}

\keywords{gravitational lensing --- intergalactic medium --- quasars: absorption lines}

\section{Introduction
\label{sec:intro}}

\ion{Mg}{2} absorbers towards the quasar sight lines have been 
systematically studied since \citet{survey} (for more recent
studies, see \citealt{sdss} and references therein). 
Similar \ion{Mg}{2} absorbers were subsequently seen in 
gamma-ray burst (GRB) spectra. \citet{grb} 
 compared GRB and quasar sight lines, and found a significantly higher incidence toward 
the former. They proposed three possible effects to explain
this discrepancy: (1) faint quasars are obscured by dust associated with the
absorbers; (2) \ion{Mg}{2}  absorbers are intrinsic to GRBs; (3) gravitational 
lensing of the GRB by the absorbers. However, they concluded that none of these 
effects provide a satisfactory explanation. 

\citet{frank} proposed a simple geometric solution to the puzzle. They
argued that if \ion{Mg}{2} absorption systems are fragmented on scales
$\lesssim 10^{16} \cm$, similar to the beam sizes of GRBs, then the
observed difference in incidence of \ion{Mg}{2} absorbers would simply
reflect the difference in the average beam sizes between GRBs and
quasars, with quasars being on average several times as big. This
explanation predicts that absorption features due to intervening
\ion{Mg}{2} cloud fragments should evolve as the size of GRB afterglow
changes, which has now been observed by \citet{hao}. However,
structures of the \ion{Mg}{2} absorbers down to the size of $\sim
10^{16}\cm$ cannot be directly inferred from their spectral
features. \citet{rauch4} put the strongest upper limits on \ion{Mg}{2}
absorber sizes to date. They observed the spectra of three images of
Q2237+0305 \citep{huchra} and found that each line of sight contained
individual \ion{Mg}{2} absorbers at approximately the same redshift,
but with distinct spectral features. Thus these absorbers are part of
a complex that extends at least $\sim 500 \pc$, but the sizes of the
individual ``cloudlets'' must be smaller than $200 - 300 \pc$ based on
the separation of the macro-images.

Some, if not all, strongly lensed quasars are also gravitationally 
microlensed by the compact stellar-mass objects in the lensing galaxy 
\citep{microlens}, and Q2237+0305 was the first lensed quasar to be found 
to exhibit significant microlensing variability \citep{micro, ogle}. The macro-image 
of a microlensed quasar is split into many micro-images, and when the 
source moves over the caustic networks induced by the microlenses, those 
micro-images 
will expand, shrink, appear, disappear and experience drastic astrometric 
shifts over timescales of months or years \citep{astrometric}. 
The angular 
sizes of major micro-images are usually of the same order as those of the 
quasars, and during the shape and position changes of these images, absorption structures
of similar scale along their sight lines will likely imprint significant 
variations on the spectrum.

\citet{lyman} pioneered the theoretical investigation of quasar microlensing
as a probe of the sub-parsec structure of intergalactic absorption systems. 
They concluded that variation in the strength of the absorption lines 
over timescales of years or decades caused by microlensing can be used 
to probe the structures of Lyman $\alpha$ clouds and associated metal-line 
absorption systems on scales $\lesssim 0.1 \pc$. 
However, as I will show, they significantly underestimated the relevant timescales for 
spectral variability given the sizes of the systems they considered. Thus, they
substantially overestimated the scales of absorption structures that 
microlensing can effectively probe.

In the following section, I lay out the basic theoretical framework
of the method. 
Then in \S~\ref{sec:huchra}, I present a numerical simulation 
of the microlensing of Q2237+0305. I show that micro-images of this 
quasar can be used to probe structures of \ion{Mg}{2} and other metal-line
absorption clouds  
on scales of $\sim 10^{14} - 10^{16}\cm$ 
by monitoring the spectral variations of absorption lines over months or years. 
Finally in \S~\ref{sec:discussion}, I summarize the results and discuss their
implications.

\section{Varying Microlensed Quasar Image as A ``Ruler''
\label{sec:theory}}

I begin with a brief summary of notation.
Subscripts ``$l$'', ``$s$'', ``$o$'' and ``$c$'' refer to the lens, source, 
observer and absorption cloud plane, respectively. 
The superscript ``$\ray$'' is used to refer to the light ray on 
the cloud plane to distinguish it from the cloud.
The angular diameter distance between object $x$ and $y$
is denoted $D_{xy}$ and is always positive regardless of which is closer;
in particular, $D_x$ refers to the angular diameter distance between 
the observer and object $x$.
The vector angular position of object $x$ is denoted 
$\vectheta_x$, while its redshift is denoted $z_x$.

Consider an absorbing cloud that is confronted with a ``bundle of
light rays'' making their way from the source to the lens to the observer.
Let ${\bf r}^{ray}_c$ be the position vector of a point on the plane of the cloud.
The line depth $\langle A_{\lambda} \rangle$ of an absorption line centered 
at wavelength $\lambda$ is given by:

\begin{equation}
\langle A_{\lambda} \rangle = {{\int\sigma_{\lambda}({\bf r}^{ray}_c) A_{\lambda}({\bf r}^{ray}_c){d^2}r^{ray}_c}{\bigg/}{\int\sigma_{\lambda}({\bf r}^{ray}_c){d^2}r^{ray}_c}}
\label{eqn:depth}
\end{equation}
where $\sigma_{\lambda}(\vecr^{ray}_c)$ is the surface density of the 
``ray bundles'' on the plane of the absorption cloud with the rays 
weighted by the surface brightness profile of the quasar 
and $A_{\lambda}(\vecr^{ray}_c)$ is the absorption fraction 
at $\vecr^{ray}_c$ for light of wavelength $\lambda$ \citep{lyman}. 

\subsection{Basic Geometric Configurations and Motions
\label{sec:basic}}

The absorption cloud can be located either between the lens and observer 
or between the lens and source. In the former case, the angular position of 
the light ray at the cloud plane is 
$\vectheta^{\ray}_c = \vectheta_i$, so $\vecr^{\ray}_c = \vectheta_{i} D_{c}$.
The projected light rays on the cloud plane maintain the exact shapes of the 
quasar images, and their physical extents are proportional to the distance to 
the observer.

If the cloud is between the lens and the source, it can be easily shown that 
$\vectheta^{ray}_c$ is given by \citep{lyman}:

\begin{equation}
\vectheta^{ray}_c = \left(1-{D_{lc} D_{s}\over D_{ls} D_c}\right) \vectheta_{i} + 
{D_{lc} D_{s}\over D_{ls} D_c} \vectheta_{s},
\label{eqn:thetacloud}
\end{equation}
so
\begin{equation}
\vecr^{ray}_c = \left(D_c -{D_{lc} \over D_{ls}}D_s\right) \vectheta_{i} + 
{D_{lc} \over D_{ls}} \vecr_{s}.
\label{eqn:rcloud}
\end{equation}
The lens-source relative proper motion, with time as measured by the observer,  
is given (in other notation) by \citet{kayser},
\begin{equation}
\vecmu_{ls} = {1 \over 1+z_s} {\vecv_{s} \over D_s} - {1 \over 1+z_l} {\vecv_{l} \over D_l} + {1 \over 1+z_l} {\vecv_{o} D_{ls} \over D_l D_s} 
\label{eqn:veff}
\end{equation}
where $\vecv_{s}$, $\vecv_{l}$, $\vecv_{o}$ are the transverse velocities of the 
source, lens and observer, relative to the cosmic microwave background (CMB).
In particular, 
\begin{equation}
  \vecv_o=\vecv_{\rm CMB}-(\vecv_{\rm CMB} \cdot \hat{\bf z})\hat{\bf z},
\end{equation}
where $\hat{\bf z}$ is the unit vector in the direction of the lens and
$\vecv_{\rm CMB}$ is the heliocentric CMB dipole velocity \citep{kocha}.

The formulae in this subject can be greatly simplified by proper choice of 
notation. To this end, I defined the ``absolute'' proper motion of an object
$x$ moving at transverse velocity $\vecv_x$ to be:
\begin{equation}
\vecmu_{\abs,x} = {1 \over 1+z_x} {\vecv_{x} \over D_x},
\label{eqn:absolute}
\end{equation}
I also define the ``reflex proper motion'' of an object $x$ relative to the
observer-lens axis to be:
\begin{equation}
\vecmu_{o,l,x} = {\rm sgn}(z_x - z_l) {1 \over 1+z_l} {\vecv_{o} D_{lx} \over D_l D_x},
\label{eqn:reflex}
\end{equation}
Then equation~(\ref{eqn:veff}) can then be simplified,
\begin{equation}
\vecmu_{ls} = \vecmu_{\abs,s} - \vecmu_{\abs,l} + \vecmu_{o,l,s}.
\label{eqn:vls}
\end{equation}
Similarly, the lens-cloud relative proper motion is given by:
\begin{equation}
\vecmu_{lc} = \vecmu_{\abs,c} - \vecmu_{\abs,l} + \vecmu_{o,l,c}
\label{eqn:vlc}
\end{equation}
Note that the last term has a different sign depending on whether 
the cloud is farther or closer than the lens (see eq.~\ref{eqn:reflex}).

\subsection{Bulk Motion of the Un-microlensed ``Ray Bundles''
\label{sec:bulk}}
Equations~(\ref{eqn:thetacloud}), (\ref{eqn:vls}) and (\ref{eqn:vlc}) are the key 
formula needed to carry out the simulation in \S~\ref{sec:huchra}. In this section,
I discuss the bulk motion and relevant timescales of the macro-image 
and its associated ``ray bundles'' on the cloud plane for the 
underlying case that the image is not perturbed by microlenses.

When the source moves at $\vecmu_{ls}$, the angular positions of the rays that
compose the images are also changing with 
respect to the observer-lens axis. If the macro-image is unperturbed by the 
microlenses, the relative proper motion $\vecmu_{li}$ is simply given by \citep{proper},
\begin{equation}
\vecmu_{li} = {\mathcal M} \cdot \vecmu_{ls},
\label{eqn:vli}
\end{equation}
where ${\mathcal M}$ is the magnification tensor.
At the same time, the intersection of ``ray bundles'' with the cloud plane
also change their angular positions relative to the lens. 
If the cloud plane is between the lens and 
the source, then from equation~(\ref{eqn:thetacloud}), the ``ray bundle''-lens
 relative proper motion $\vecmu_{lc}^{\ray}$ is a linear combination 
of $\vecmu_{ls}$ and $\vecmu_{li}$ weighted by distances,
\begin{equation}
\vecmu_{lc}^{\ray} = \left( 1-{D_{lc} D_{s}\over D_{ls} D_c} \right) \vecmu_{li} + 
{D_{lc} D_{s}\over D_{ls} D_c} \vecmu_{ls}.
\label{eqn:vlcray}
\end{equation}
Substituting equation~(\ref{eqn:vli}) into equation~(\ref{eqn:vlcray}) yields
the bulk proper motion of the un-microlensed ``ray bundle'' relative to the lens,
\begin{equation}
\vecmu_{lc,bulk}^{\ray} = 
\left[ 
{\left( 
{1-{D_{lc} D_{s}\over D_{ls} D_c}}
\right) 
{\mathcal M} + {D_{lc} D_{s}\over D_{ls} D_c}{\mathcal I}}
\right] 
\cdot \vecmu_{ls}.
\label{eqn:vlcbulk}
\end{equation}
where $\mathcal I$ is the unit tensor.

When the ``ray bundles'' are between the source and the lens, their relative 
bulk proper motion is simply,
\begin{equation}
\vecmu_{lc,bulk}^{\ray} = \vecmu_{li} = {\mathcal M}(\vectheta_i) \cdot \vecmu_{ls}.
\label{eqn:vlcbulk2}
\end{equation}

There are two major effects that may induce variability of absorption lines 
toward a lensed quasar. One is the creation, destruction, distortion and astrometric 
shifts of micro-images, which probe structures similar to the size of the 
micro-images. The other is the bulk motion of the ``ray bundles'' relative
to the cloud. On the cloud plane, this motion has an angular speed of 
$\Delta{\vecmu_{lc,bulk}} = \vecmu_{lc,bulk}^{\ray} - \vecmu_{lc}$, 
and the time $t_{cc}$ required for the ``ray bundles'' to cross a cloud of transverse 
size $R_c$ is,
\begin{equation}
t_{cc} = {R_c \over D_c |\Delta{\vecmu_{bulk,lc}}|} 
= {R_c \over D_c |\vecmu_{bulk,lc}^{\ray} - \vecmu_{lc}|}.
\label{eqn:tcc}
\end{equation} 

\section{Application to Q2237+0305
\label{sec:huchra}}

In October 1998, \citet{rauch4} obtained high-resolution Keck spectra of 
images A, B and C of Q2237+0305. They found \ion{Mg}{2} 
absorption lines at redshifts of 
$z = 0.5656$ and $z = 0.827$ in the spectra of all three images, but absorption
profiles of the individual sight lines differed (e.g., Figs.~6 and 10 of their 
paper). Therefore, they concluded that the 
\ion{Mg}{2} complexes giving rise to these absorption features must be larger than 
$\sim 0.5 \kpc$, while the individual \ion{Mg}{2}  components must be smaller
than $\sim 200 - 300 h_{50}^{-1} \pc$. Q2237+0305 is also one of the most
observed and studied lensed quasars with obvious microlensing features, and the properties of the system are well known. These factors make it
an ideal object to investigate.

The comprehensive statistical study of this lens by \citet{kocha} showed that
the size of the quasar is $\sim 10^{15} h^{-1} \cm - 10^{16} h^{-1} \cm$. 
\citet{size} demonstrated that the source size has a much more significant
effect on microlensing models than the source brightness profile. For simplicity, 
in the simulation, I model the source as uniform disks with four different sizes: 
$10^{15} h^{-1} \cm$, $3\times10^{15} h^{-1} \cm$, $5\times10^{15} h^{-1} \cm$
and $10^{16} h^{-1} \cm$. Uniform grids of rays are traced from image plane to
source plane \citep{thesis}. Because structures of interest have similar 
sizes as the source, finite-source effects must be taken into account. The 
grid size used has an angular scale $1/10$ of the smallest source. \citet{kocha}
demonstrated that a Salpeter mass function cannot be distinguished from a uniform mass
distribution and found the mean stellar mass to be 
$\langle{\it M}\rangle \sim 0.037 h^{-1} M_\odot$. For simplicity, 
I assign all stars in the simulation the same mass of $0.04 h^{-1} M_\odot$. 
Rays are shot from a region extending $47 \langle \theta_{\e} \rangle$ on each side, 
where $\langle\theta_{\e}\rangle$ is the Einstein radius of a $0.04 h^{-1} M_\odot$ star.
I adopt a convergence and shear for image A of $(\kappa, \gamma) = (0.394, 0.395)$
from \citet{kocha}, and set the stellar surface density $\kappa_* = \kappa$. 
Four different trajectories, oriented at $0, 30, 60$ and $90$ degrees
with respect to the direction of the shear are studied. Positions on both the image 
plane and the source plane are recorded once a ray falls within a distance of $2$ 
times the largest source size from any trajectory on the source plane. A total length
of $5 \langle\theta_{\e}\rangle$ along each trajectory is considered.

Throughout the paper, I adopt a $\Lambda$CDM cosmology with ${\Omega}_{m} = 0.3$, ${\Omega}_{\Lambda} = 0.7$ and ${\rm H}_0 = 100 \,h\,\kms
\,{\rm Mpc}^{-1}$.
The lens and source are at redshifts $z_l = 0.0394$ and $z_s = 1.695$ \citep{huchra}. 
These imply $(D_s, D_l, D_{ls}) = (1223, 113, 1180) \, h^{-1} \, \mpc$.
Based on \citet{kocha}, I adopt transverse velocities of the lens, source and 
observer of $({\rm v}_l, {\rm v}_{s}, {\rm v}_{o}) = (300, 140, 62) \kms$. 
The lens and source absolute proper 
motions and the source reflex proper motion are 
$(\mu_{\abs,l}, \mu_{\abs,s}, \mu_{o,l,s}) = (0.54, 0.009, 0.11) \,h\,\muas\,
{\rm yr}^{-1}$. 
So the lens absolute proper motion dominates the lens-source relative proper motion. 
It can easily be shown that the absolute and reflex proper motions of the 
cloud are much smaller than the absolute proper motion of the lens unless 
 the cloud redshift is close to or smaller than the lens redshift. 
For Q2237+0305, the lens redshift is very small compared to the source, 
so most likely $z_c \gg z_l$. Thus in following analysis, I focus on cloud 
with redshift $z_c > z_l$, and ignore absolute and reflex proper motions of 
the source and the cloud. In this case, the source and the cloud share the same
relative proper motion,
\begin{equation}
\vecmu_{lc} = \vecmu_{ls} = - \vecmu_{\abs,l}
\label{eqn:nopeculiar}
\end{equation}

In the simulation, as a practical matter, I hold the positions of the observer,
lens galaxy (as well as its microlensing star field) fixed, and allow the source
to move through the source plane at $\vecmu_{ls}$. Then by equation~(\ref{eqn:nopeculiar}),
the cloud has the same relative proper motion as the source. At any given time, the
angular positions of the ``ray bundles'' are calculated using 
equation~(\ref{eqn:thetacloud}). Then by simply subtracting the angular position of the
source at that time, the ``ray bundles'' positions are transformed to the reference 
frame of the cloud.

During short timescales, microlensing causes centroid shifts of the 
macro-image \citep{shift,astrometric}, 
with respect to the steady bulk motion of ``ray bundles'' relative
to the cloud, which is described in \S~\ref{sec:bulk}. 
If the direction of relative lens-source proper motion is
the same as the lens shear, then by subtracting $\vecmu_{ls}$ 
from equation~(\ref{eqn:vlcbulk}), one finds that the ``ray bundles'' have their 
maximum bulk proper motion relative to the cloud,

\begin{eqnarray}
\Delta{\mu_{bulk,lc,{\rm max}}} 
& &= \left( 1-{D_{lc} D_{s} \over D_{ls} D_c} \right) \left({1 \over 1-\kappa-\gamma} -1 \right) \mu_{ls} \nonumber \\
& & \sim 2 \left(1- 1.037 {D_{lc} \over D_c}\right) \,h\,\muas\,{\rm yr}^{-1}.
\label{eqn:veffcloud2}
\end{eqnarray}
If the source moves perpendicular relative to 
the lens shear, $\Delta{\mu_{lc}} = 
[1-{D_{LC} D_{S}/ (D_{LS} D_C)}]|1/(1-\kappa+\gamma) -1| \mu_{ls}$, which is
approximately 0 for the $(\kappa, \gamma)$ of image A.

\begin{figure}
\epsscale{0.8}
\plotone{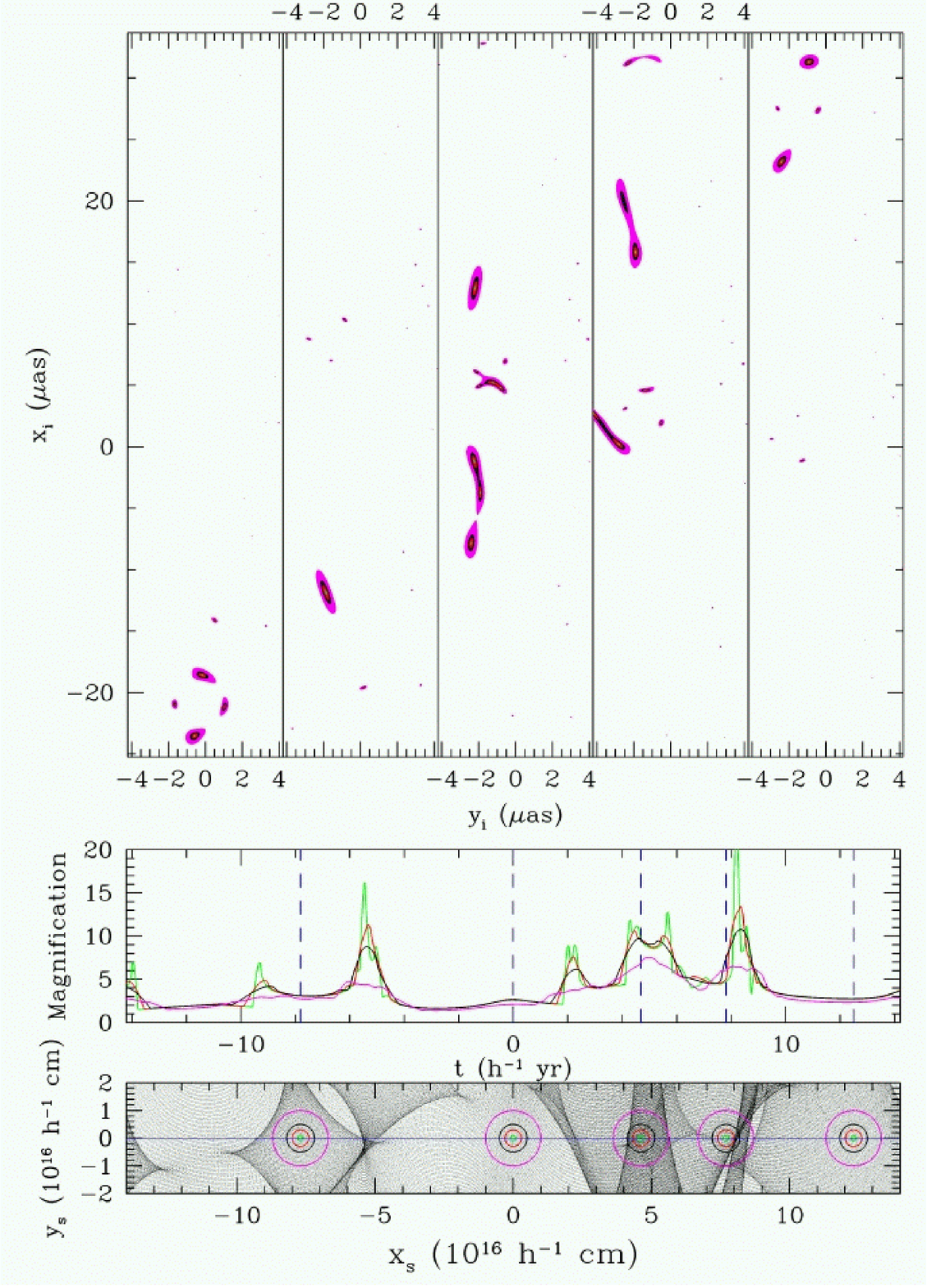}
\caption{\label{fig:lc}
Caustics network (bottom panel), light curve (middle panel)
and images relative to the source position (top panel) 
for a source trajectory that is parallel 
to the lens shear (the direction of x-axis). Different colors 
represent the 4 different source sizes.
}
\end{figure}

\begin{figure}
\epsscale{0.8}
\plotone{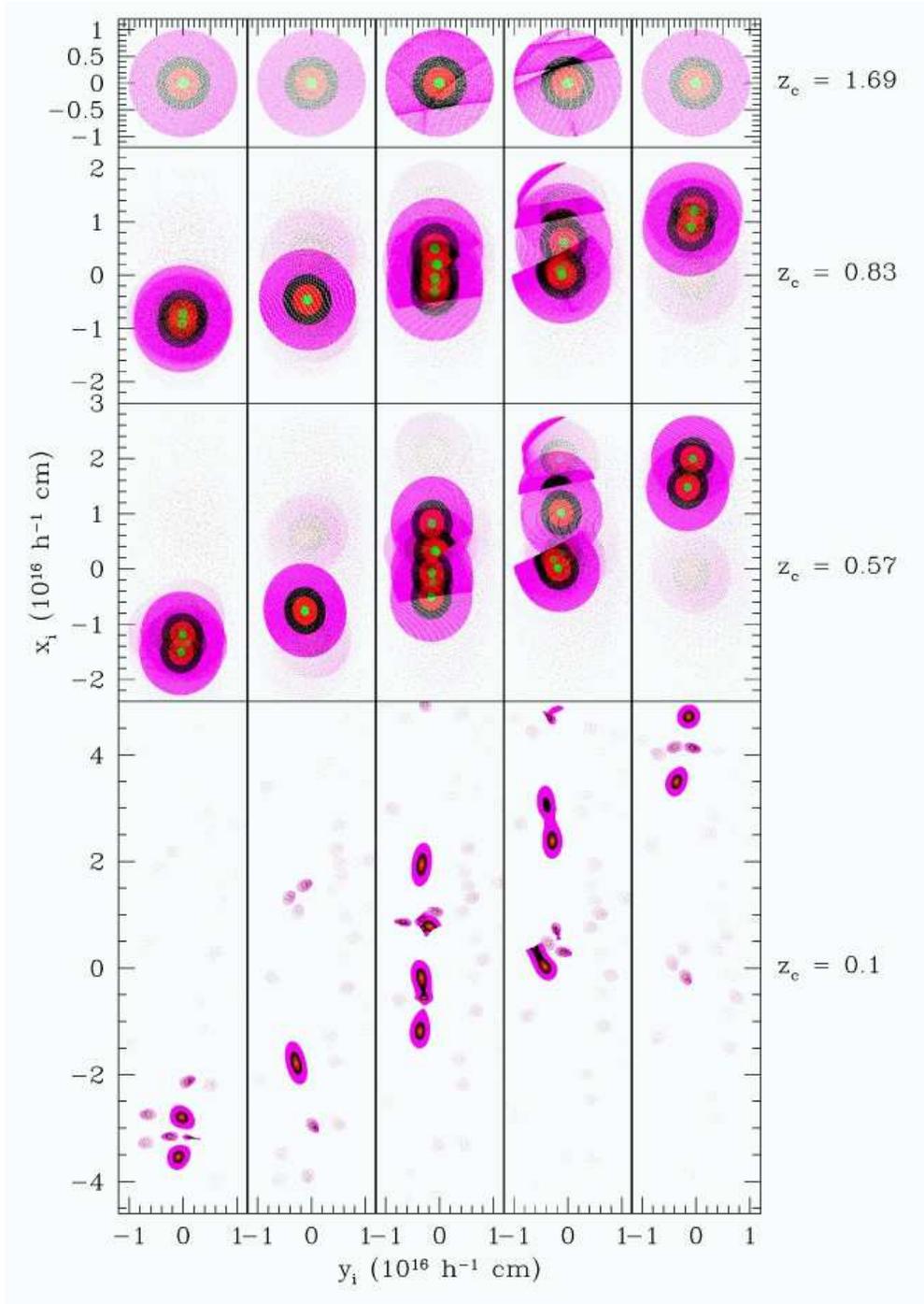}
\caption{\label{fig:parallel}
Physical positions of ``ray bundles'' in the frame of cloud 
at redshifts $0.1$, $0.57$, $0.83$ and $1.69$. 
The 5 different columns correspond to the source positions 
shown in Figure~\ref{fig:lc}. x-axis is the direction of the
lens shear.
}
\end{figure}

Figure~\ref{fig:lc} and Figure~\ref{fig:parallel} show the 
results for a source trajectory that is parallel to the 
shear direction. The bottom panel of Figure~\ref{fig:lc} 
shows the magnification pattern on the source plane, 
with a series of 4 concentric circles centered
at 5 source positions; and the top panels show the images at
these positions relative to the source (which has the same
proper motion as the cloud). 
Different colors represent different source sizes. The middle panel of 
Figure~\ref{fig:lc} shows 
the light curves for the 4 source sizes with the blue
dash lines used to mark the times for the five source positions.
Figure~\ref{fig:parallel} shows the  ``ray bundles'' positions 
in the cloud frame at redshifts $1.69$, $0.83$, $0.57$, $0.1$.
The 5 different columns show the 5 positions corresponding to 
those in Figure~\ref{fig:lc}. 

In the top row of Figure~\ref{fig:parallel}, one can see that 
the ``ray bundles'' at $z_c = 1.69$, which 
is very close to the quasar, have almost exactly 
the same size and shape as the source and that the bundles show almost no bulk 
motion. The density 
of ``ray bundles'' clearly have the imprints from magnification pattern 
shown in the bottom panel of Figure~\ref{fig:lc}. So if an absorption cloudlet
 has a similar or somewhat
smaller size than a source that is sitting directly behind it, the
depth of its corresponding absorption line will change dramatically as the 
source crosses the caustics. The magnification close to a fold caustic is 
proportional to the inverse square root of the distance from it, 
so for a cloud with angular size $\theta_c < \theta_s$, the fractional change 
in absorption line depth caused by the caustics scales as 
$(\theta_c/\theta_s)^{3/4}$. Hence, for cloud close to the
quasar redshift, structures on scale $\sim 10^{14} - 10^{16} h^{-1} \cm$ (depending on the source size) will be probed 
over few-month to few-year timescales (i.e., 
the timescale of typical caustics crossings).

The fourth row of panels of Figure~\ref{fig:parallel}
shows ``ray bundles'' for $z_c = 0.1$, 
which is close to the lens redshift. A distinct difference between these 
``ray bundles'' from the ones at $z_c = 1.69$ is that they are split 
into many groups of
bundles, which correspond to the micro-images in the top panel
of Figure~\ref{fig:lc}. I dub these
groups of ``ray bundles'' as micro-images in the following discussions. Most of these 
micro-images are stretched one-dimensionally, and most rays 
are concentrated in a few {\it major} micro-images. The micro-images in 
different columns have drastically different morphologies and positions. If there
are cloudlets of similar sizes as these micro-images distributed on the
cloud plane, then the absorption spectra will show multi-component absorption
features at any given time. These components will experience
drastic changes in line depth, with some components disappearing and other new
components appearing during the course of months or years, as the
source crosses the microlensing caustics. Another important characteristic is
that the bulk of these micro-images are moving in the same direction as the
source. This motion is described by equation~(\ref{eqn:veffcloud2}), 
which yields $\sim 0.8 \,h\,\muas\,
{\rm yr}$. So from equation~(\ref{eqn:tcc}), structures as large as 
$\sim 3.0 n \times 10^{16} \,h^{-1} \,\cm$  will be 
crossed in $n\times 10 \,h^{-1}\, {\rm yr}$ by the bulk motion of the ``ray bundles''. Hence, the
effects caused by the micro-images and the bulk motion of the bundles together
probe scales of $\sim 10^{14} - 10^{16} h^{-1} \cm$ on timescales of months to years.

The second and third row of panels in 
Figure~\ref{fig:parallel} refer to clouds at intermediate redshifts 
between the lens and the source. Their redshifts, $z_c = 0.83$ and $z_c = 0.57$,
 are close to the \ion{Mg}{2} absorption systems observed by \citet{rauch4}. As expected,
the characteristics of these ``ray bundles'' are intermediate to those
shown in rows 1 and 4. The overall shapes of the micro-images are close to 
that of the source, with magnification patterns imprinted on them. And they also clearly
show multiple components, which appear and disappear as the source crosses
caustics. The angular sizes of the images are close to that of the (unmagnified) quasar. 
These micro-images could probe clouds with angular size from a factor of few smaller 
to a factor of few larger than the source size, which corresponds to 
scales of $\sim 10^{14} - 10^{16} h^{-1} \cm$. From equation~(\ref{eqn:tcc}) 
and (\ref{eqn:veffcloud2}), the bulk motion of the bundles will probe cloud structures
$\sim 0.8 n \times 10^{16} \,h^{-1}\,\cm$ and $\sim 1.3 n \times 10^{16}\,h^{-1}\,\cm$ during $n\times 10 \,h^{-1}\, {\rm yr}$ 
for $z_c = 0.83$ and $z_c = 0.57$, respectively. 

For a source trajectory that is perpendicular to the shear, there will be almost no 
bulk motion relative to the cloud for the ``ray bundles'', while the magnitude of 
bulk motions for intermediate trajectories is a fraction of the parallel case 
depending on their angles relative to the shear direction. And the micro-images
of these trajectories share similar properties with those of the trajectory that 
is parallel to the shear. Figure~\ref{fig:parallel} shows that micro-image motions
are in the same orders as the bulk motion of macro-image over the timescales considered. 
Therefore, trajectory direction only has a modest impact on the cloud sizes probed. 

These results are in great contrast with those of \citet{lyman}, who claimed 
quasar microlensing for image A of Q2237+0305 can induce considerable variability 
of absorption lines 
associated with structures as large as $0.1 \pc$ during the course of years to 
decades. According to their analysis, the effect is largest when the 
cloud is very close to the source, and for example, the timescale of line strength 
variation for a $0.1 \pc$ cloud very close to the source is given as $\sim 16.2 
{\rm yr}$. 
However, in their analysis, they effectively assumed the relative lens-cloud
proper motion $\mu_{lc} = 0$. This would lead to a timescale 
$t_{cc} \sim R_c D_l/ ({\rm v}_l D_s) \sim 16 (R_c/0.1 {\pc}) {\rm yr}$ for 
absorption cloud near the source redshift 
(they adopted ${\rm v}_l = 600 \kms$), which
is in agreement with column 4 of their Table 1. In fact, I showed that,
when peculiar velocities of the source and cloud are ignored,  
$\vecmu_{lc} = \vecmu_{ls}$ (eq.~{\ref{eqn:nopeculiar}}), which is not negligible.
Even considering realistic peculiar motions of the source and the cloud, it still 
leads to time scales that are more than one order of magnitude slower than those
predicted for $\mu_{lc} = 0$. In addition, 
when the cloud is close to the source,
the angular sizes of the clouds they considered are orders of magnitudes larger
than their source size, so effects of changes in magnification pattern on the source
plane alone have very little impact as well.
Therefore, \citet{lyman} significantly overestimated the cloud size to which microlensing 
is effectively sensitive.

\section{Discussion and Conclusion
\label{sec:discussion}}

I have shown that there are two effects that might induce variation
of absorption lines along the sight lines to lensed quasar. One effect is 
caused by the drastic morphological and positional changes of micro-images when
the source crosses the caustic network. The other effect is due to the bulk motion
of the ``ray bundles'' relative to the absorption clouds. I have laid out a
basic framework in studying these effects for microlensed quasars in general.
And in particular, I perform numerical simulations to apply the method to 
image A of Q2237+0305. I demonstrated that the combinations of these two 
effects probe $10^{14} - 10^{16} \,h^{-1}\, \cm$ absorption cloudlets between the lens and 
the source over timescales of months to years. The existence of these cloudlets will
be revealed by either changes in line depths or appearances/disappearances of 
multi-absorption components. Spectra should preferably be taken during the course of
caustic crossings, which can be inferred from photometric monitoring programs of
lensed quasars. In fact, the \ion{Mg}{2} lines observed by \citet{rauch4} about 8
 years ago already show different multi-components along sight lines of three different
macro-images, implying they might be caused by fragmented cloudlets with similar
sizes as the micro-images. A similar high-resolution spectrum taken in the near future
would provide a definitive test of the existence for structures of \ion{Mg}{2} or other 
metal-line absorbers at the scales of $10^{14} - 10^{16} \,h^{-1}\, \cm$. If the spectral variations
are indeed observed, a statistical study similar to \citet{kocha} will be required to
infer the properties of the cloudlets. Moreover, a time series of spectra 
may provide additional constraints to quasar microlensing models.

\acknowledgments

I would like to thank Kris Stanek for inspiring discussions that
motivated this project. I am especially grateful to Andy Gould for his
thorough discussions and useful advice during the completion of this
work.  I thank Chris Kochanek for his help and insightful comments.
Support for this work was provided by NSF grant AST-042758.

\end{document}